\documentstyle[prl,aps,epsf]{revtex}
\begin{document}
\draft
\twocolumn[
\hsize\textwidth\columnwidth\hsize\csname@twocolumnfalse\endcsname
\preprint{}
\title{Local Quasiparticle States around an Anderson Impurity in a 
$d$-Wave Superconductor: Kondo Effects} 
\author{Jian-Xin Zhu and C. S. Ting}
\address{ 
Texas Center for Superconductivity and Department of Physics, 
University of Houston, Houston, Texas 77204
}
\maketitle
\begin{abstract}
The mutual influence between a Kondo impurity and a $d$-wave 
superconductor is numerically studied within the slave-boson mean-field 
approach. The Bogoliubov-de Gennes type equations are derived and 
solved using exact diagonalization. We show that a critical coupling strength,
above which the Kondo effect takes place, exists regardless of whether the band 
particle-hole symmetry is present or not. In the Kondo regime the 
double resonant quasiparticle peaks are found in the local density of 
states (LDOS) both directly at the 
impurity and around its neighbors, which is in sharp contrast to the case of 
nonmagnetic unitary impurities, where the LDOS vanishes at the 
impurity site.  
\end{abstract} 
]

\narrowtext

The identification of $d$-wave pairing state in high-$T_c$ 
superconductors~\cite{Harl95} 
has renewed much interest in low-energy quasiparticle states in 
unconventional 
superconductors. Different from conventional $s$-wave superconductors 
having a constant energy gap, the $d$-wave 
superconductors have an anisotropic energy gap which changes sign along 
the nodal directions on the essentially cylindrical Fermi surface. The 
anisotropy of the energy gap makes the $d$-wave superconductors very 
sensitive to impurity or other potential scattering, in contrast to the 
$s$-wave superconductors which, as dictated by the Anderson 
theorem~\cite{Ande59}, 
are almost unaffected by the presence of nonmagnetic impurities. As an 
important signature of the sign change of $d$-wave order parameter (OP), 
Balatsky, Salkola 
and co-workers~\cite{Bala95,Salk96}, based on the T-matrix approximation, 
theoretically predicted that a strong atomic-like nonmagnetic impurity 
can induce a virtual bound state. Later studies~\cite{Tsuc99,Zhu00a} 
showed that the 
resonant state would show up in the local density of states (LDOS), which 
could be measured by the local differential tunneling conductance. 
Recent scanning 
tunneling spectroscopy measured on high-$T_c$ superconductor 
Bi$_2$Sr$_2$CaCu$_2$O$_{8+\delta}$ (BSCCO) with individual atomic-scale 
defects~\cite{Yazd99} or strong impurities~\cite{Huds99,Pan00}, has 
provided 
strong evidence for the existence of low-energy quasi-bound states. 
Especially, the experiment by Pan {\em et al.}~\cite{Pan00} on the 
observation 
of quasiparticle resonant states has a high spatial and energy resolution 
and the identity of impurity atoms is known in a controlled way. Very 
interestingly, the imaging of the effect of a single zinc (Zn) impurity 
atom showed the strongest intensity of the LDOS 
at resonance directly on the Zn atom, which is opposite to the earlier 
theories~\cite{Bala95,Salk96,Tsuc99,Zhu00a}. In our previous 
work~\cite{Zhu00b}, we explained the novel 
pattern of the tunneling conductance in terms of the blocking effect of 
the BiO and SrO layers which exist between the tunneling tip and the 
CuO$_2$ layer being probed. This explanation may not exclude other 
possibilities and needs further experimental justification~\cite{Note1}. 
Although almost  all existing theories are treating the Zn atom as a static 
potential scatterer, there exists a few experiments~\cite{Juli00} indicating 
that Zn impurity may possess a magnetic moment. 
It is now an appropriate time to consider the effects of 
dynamic impurities on the quasiparticle properties of $d$-wave 
superconductors. The study of a dynamic impurity  in a superconductor 
began with the works of Yu~\cite{Yu65} and Shiba~\cite{Shib68} on magnetic 
impurities in $s$-wave 
systems. The solving of this problem is quite challenging mainly 
due to the difficulty in treating the dynamical correlations of the 
coupled impurity-conduction electron system with pair correlation. 
The problem of Kondo impurities in a 
$d$-wave system has been studied by focusing on either the moment 
formation~\cite{Simo99,Salk97}  
or the suppression of superconducting transition 
temperature by them~\cite{Bork94}. Quite recently, within a Kondo spin model, 
Tsuchiura {\em et al.}~\cite{Tsuc00} have studied the LDOS around a magnetic 
impurity by treating the impurity spin as an effective local magnetic 
field and the screening of the impurity spin by the quasiparticles was 
not considered there. However, other new theories~\cite{Naga97,Cass97} 
argued that the moment may be screened deep in a $d$-wave  
superconducting state. 
Experimentally, the specific heat measurement of Zn-doped 
YBa$_2$Cu$_3$O$_{6.95}$~\cite{Siss00} provided a possible evidence of 
Kondo screening in the superconducting state.

In this paper we study the Kondo effects of an Anderson impurity  coupled 
with the conduction electrons of a $d$-wave superconductor. Within a 
slave-boson mean-field approach, we are able to exactly solve the 
composite system with the self-consistency for the $d$-wave OP included. 
It is shown that a critical coupling strength, above which the Kondo 
effect emerges, exists regardless of the band particle-hole (PH) symmetry 
being present or not.  We also show in the Kondo regime that  
a double resonant-peak structure shows up in the LDOS
both at the impurity 
and at its nearest neighboring sites. This result is different from those 
for the static impurities that the LDOS at the impurity site 
exhibits only a single peak below (or above) the Fermi energy for a weak 
repulsive (or attractive) scattering center while vanishes in the unitary 
scattering limit. 

To describe the impurity and its interaction with the band electrons of 
the $d$-wave superconductor defined on a two-dimensional lattice, 
we start with an Anderson-like Hamiltonian, 
which consists of three parts: 
$H=H_{S}+H_{imp}+H_{S\mbox{-}imp}$. Here 
$H_{S}=-\sum_{{\bf ij},\sigma} (t_{\bf ij}+\mu\delta_{\bf ij}) c_{{\bf 
i}\sigma}^{\dagger} c_{{\bf j}\sigma} 
+\sum_{\bf ij} (\Delta_{\bf ij} c_{{\bf 
i}\uparrow}^{\dagger}c_{{\bf j}\downarrow}^{\dagger}+\Delta_{\bf ij}^{*} 
c_{{\bf j}\downarrow}c_{{\bf i}\uparrow}) 
+\sum_{{\bf i}\neq {\bf j}}\frac{\vert \Delta_{\bf ij} \vert^{2}}{g_{\bf ij}}
$, is the Hamiltonian for the host $d$-wave superconductor, where $c_{{\bf 
i}\sigma}^{\dagger}$ ($c_{{\bf i}\sigma}$) are the 
creation (annihilation) operators of a conduction electron with spin 
$\sigma$ at the ${\bf i}$th site, $t_{\bf ij}$ is the hopping integral 
between two nearest neighbor sites, $\mu$ is the chemical potential, 
$g_{\bf ij}$ represents the strength of $d$-wave pairing interaction, 
and $\Delta_{\bf ij}$ is the bond pairing amplitude which is in turn 
determined self-consistently as $\Delta_{\bf ij}=g_{\bf ij}\langle 
c_{{\bf i}\uparrow}c_{{\bf j}\downarrow}\rangle$;
$H_{imp}=\sum_{\sigma} \epsilon_d d_{\sigma}^{\dagger}d_{\sigma} 
+U n_{d\uparrow}n_{d\downarrow}$
is the Hamiltonian for the Anderson impurity, where 
$d_{\sigma}^{\dagger}$ ($d_{\sigma}$) is the creation (annihilation) 
operator of the impurity electron with spin $\sigma$, 
$n_{d\sigma}=d_{\sigma}^{\dagger}d_{\sigma}$,
$\epsilon_{d}$ 
is the bare impurity energy level with respect to the chemical potential 
of band electrons, 
and $U$ is the on-site repulsive interaction of the impurity electrons; 
and 
$H_{S\mbox{-}imp}={\cal V}\sum_{\sigma} 
(c_{{\bf i}={\bf 0}+\boldmath{\mbox{$\delta$}},\sigma}^{\dagger} 
d_{\sigma} +d_{\sigma}^{\dagger} 
c_{{\bf i}={\bf 0}+\boldmath{\mbox{$\delta$}},\sigma})$
represents the hybridization of the impurity with the conduction 
electrons, where without loss of 
generality, we have chosen the impurity to be located at the origin 
${\bf i}={\bf 0}$, $\boldmath{\mbox{$\delta$}}$ are the unit vectors of 
the square lattice, and ${\cal V}$ is the coupling strength. 
For our purpose of studying the low temperature physics, we assume that 
the on-site interaction $U$ is infinite so that the double occupancy on the 
impurity is forbidden, which allows us to apply the slave-boson 
mean-field theory~\cite{Barn76}. This approach has been applied 
successfully to the low-temperature properties of a Kondo impurity in the 
presence of normal state conduction electrons~\cite{Read83}, and to 
transport properties of a quantum dot connected to a normal and to a 
(conventional)  superconducting lead~\cite{Schw99,Cler99}, or embedded 
into a (conventional) Josephson junction~\cite{Rozh99}. 
By writing the impurity electron operator as 
$d_{\sigma}=b^{\dagger}f_{\sigma}$, where $f_{\sigma}$ and $b$ are the 
spinon and holon operators and subject to the single occupancy constraint 
$\sum_{\sigma} f_{\sigma}^{\dagger}f_{\sigma} +b^{\dagger}b=1$, in the 
mean-filed approximation, we are able to obtain 
$H_{imp}=\sum_{\sigma}(\epsilon_{d}+\lambda_{0})f_{\sigma}^{\dagger}f_{\sigma}
+\lambda_{0}(b_{0}^{2}-1)$, and $H_{S\mbox{-}imp}={\cal V}b_{0} 
\sum_{\sigma}(c_{{\bf i}={\bf 
0}+\boldmath{\mbox{$\delta$}},\sigma}^{\dagger}f_{\sigma}
+f_{\sigma}^{\dagger}c_{{\bf i}={\bf 
0}+\boldmath{\mbox{$\delta$}},\sigma})$. 
Here the holon operator $b$ has been replaced by a $c$-number $b_0$, 
$\lambda_0$ is the Lagrange multiplier introduced to enforce the single 
occupancy constraint. Then the composite Hamiltonian can be diagonalized 
by performing the canonical transformation for both the conduction and 
the impurity electrons, which leads to the Bogoliubov-de Gennes equations:
\begin{equation}
\sum_{\bf j}\left( \begin{array}{cc} 
H_{\bf ij} & \Delta_{\bf ij} \\
\Delta_{\bf ij}^{*} & -H_{\bf ij} 
\end{array} \right) 
\left( \begin{array}{c} 
u_{\bf j}^{n} \\ v_{\bf j}^{n} 
\end{array} \right)
=E_{n}
\left( \begin{array}{c} 
u_{\bf i}^{n} \\ v_{\bf i}^{n} 
\end{array} \right)\;,
\label{EQ:BdG}
\end{equation}
where $(u_{\bf i}^{n},v_{\bf i}^{n})^{\mbox{\small Transpose}}$ is the 
quasiparticle wavefunction with eigenvalue $E_{n}$,
\begin{equation}
H_{\bf ij}=-\tilde{t}_{\bf ij} -\mu \delta_{{\bf i}\neq {\bf 0},{\bf j}}
+(\epsilon_{d}+\lambda_0)\delta_{\bf i0}\delta_{\bf ij}\;,
\end{equation}
with $\tilde{t}_{\bf ij}={\cal V}b_{0}$ for $({\bf ij})
=(\boldmath{\mbox{$\delta$}}{\bf 0})$ or $({\bf 0} 
\boldmath{\mbox{$\delta$}})$ and $t$ otherwise, 
$\Delta_{\bf ij}$ and $b_0$ are subject to the self-consistent 
condition, 
\begin{equation}
\Delta_{\bf ij}=\frac{g_{\bf ij}}{2}\sum_{n}(u_{\bf i}^{n}v_{\bf j}^{n*}
+u_{\bf j}^{n} v_{\bf i}^{n*})\tanh(E_{n}/2k_{B}T)\;,
\end{equation}
and
\begin{equation}
b_{0}^{2}=1-2\sum_{n}\{ \vert u_{\bf 0}^{n}\vert^{2} f(E_{n}) 
+\vert v_{\bf 0}^{n}\vert^{2}[1-f(E_{n})]\}\;,
\label{EQ:SINGLE}
\end{equation}
with $f(E)=[1+\exp(E/k_{B}T)]^{-1}$ the Fermi distribution function.
Here we have assumed the conduction and the impurity electrons are 
of different nature and take the pairing interaction $g_{\bf ij}=0$ 
between them, and otherwise $g_{\bf ij}=g$ to be a constant.
Notice that the eigenvalues and the corresponding eigenfunction amplitudes
in Eq.~(\ref{EQ:SINGLE}) is implicitly $\lambda_0$-dependent. 
The single occupancy constraint, as given by Eq.~(\ref{EQ:SINGLE}), 
is not sufficient to determine the values of $b_0$ and $\lambda_0$. 
For this purpose, we need to  minimize the 
free energy, which is derived for the first time in terms of the 
quasiparticle wavefunction as:
\begin{eqnarray}
F&=&2k_{B}T\sum_{n}\ln [1-f(E_{n})]
-\sum_{n}\sum_{{\bf i,j}}[\Delta_{\bf ij} u_{\bf j}^{n*}v_{\bf 
i}^{n} +\mbox{c.c.}]
\nonumber \\
&&+\sum_{n}\sum_{{\bf i},{\bf j}}\{ [-\tilde{t}_{\bf ji}
+(\epsilon_{d}+\lambda_0)\delta_{\bf i0}\delta_{\bf ij}-\mu 
\delta_{{\bf i}\neq {\bf 0},{\bf j}}]v_{\bf j}^{n}v_{\bf 
i}^{n*}+\mbox{c.c.}\} \nonumber \\
&&+\sum_{\bf ij}\frac{ \vert \Delta_{\bf ij}\vert^{2} }{g_{\bf ij}}
+\lambda_{0}(b_{0}^{2}-1)\;.
\end{eqnarray}
This minimization will give two equations for $\lambda_0$ and $b_0$, 
one of which is just the single occupancy constraint
Eq.~(\ref{EQ:SINGLE}). 
When applied to other physical problems, this expression for the free 
energy is very useful in determining the stability of competing pairing 
states. 

\begin{figure}[h]
\centerline{\epsfxsize=3.0truein \epsfbox{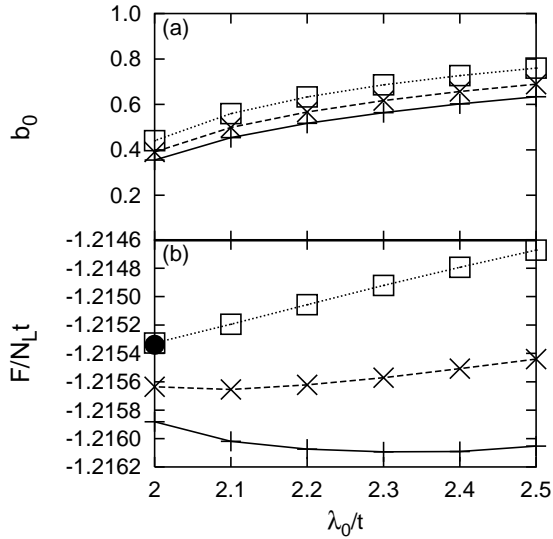}}
\vskip 1.2truein
\caption{The parameter $b_0$ (a) and the free energy per site as a 
function of the Lagrange multiplier $\lambda_0$ for  
the coupling strength ${\cal V}=1.0t$ (solid line), $0.8t$ (dashed line), 
and $0.6t$ (dotted line). Also shown in (b) the value of $F/N_{L}t$ 
with $\lambda_0=2t$ and $b_0=0$ (filled circle).
The chemical potential $\mu=-0.5t$, the bare 
impurity level $\epsilon_d=-2t$, the $d$-wave pairing interaction $g=2t$. } 
\label{FIG:FE}
\end{figure}

Equation~(\ref{EQ:BdG}) always has a trivial solution with 
$\lambda_0=-\epsilon_{d}$ and $b_0=0$, which describes the impurity 
decoupled with the conduction electrons. This trivial critical point can 
be shown not necessarily to have the lowest free energy. 
We solve Eq.~(\ref{EQ:BdG}) by taking $\lambda_0$ as a fixed parameter, and 
calculate the free energy with
the obtained value of $b_0$ and $\Delta_{\bf 
ij}$. By varying the value of $\lambda_0$ and repeating the same 
procedure, we obtain the free energy as a function of $\lambda_0$. 
Comparing the minimum free energy with that of $\lambda_0=-\epsilon_d$ 
and $b_0=0$, the true solution for $\lambda_0$ and $b_0$ is then 
determined. The numerical calculation is performed on a square lattice 
with size $N_{L}=N_x\times N_y=35\times 35$ and averaged over 36 
wavevectors in the supercell Brillouin zone. Throughout the work, we are 
interested in the zero temperature limit, and take the pairing 
interaction $g=2t$, the bare impurity level $\epsilon_d=-2t$. 
With respect to the real materials, this value of $g$ leads to
an exaggerated amplitude of the energy gap. Our choice of the parameters 
is motivated by the desire to have a short coherence length. The  calculations
with smaller values of energy gap have shown similar results and no
qualitative changes appear.                         
In Fig.~\ref{FIG:FE}, we plot $b_0$ and the free energy per site as a
function of $\lambda_0$ for various values of the coupling 
strength ${\cal V}$. Also shown is  the value of the\

\noindent
free energy $F_0$ obtained 
with $\lambda_0=-\epsilon_d$ and $b_0=0$, which is independent of ${\cal 
V}$. We find a critical value $V_{c}\approx 0.6t$ for both $\mu=-0.5$ and 
$0$, below which the 
minimum free energy is larger than $F_0$. In this regime, the impurity is 
decoupled with conduction band, and the electron on the impurity site is 
described by the spinon only. When ${\cal V}>{\cal V}_{c}$, this free local
spin state is unstable against the coupling to the 
conduction electrons. We find: For the coupling strength ${\cal 
V}=1.0t$, $0.8t$, and $0.6t$, respectively, $(\lambda_0,b_0)=(2.3t,0.56)$, 
$(2.1t,0.50)$, and $(2t,0)$ for 
$\mu=-0.5t$, which breaks the band 
PH symmetry; while 
$(\lambda_0,b_0)=(2.4t,0.59)$, $(2.1t,0.42)$, and $(2.0t,0)$ 
for $\mu=0$, which conserves the band PH symmetry. 
Our numerical result is consistent with the conclusion by Withoff and 
Fradkin~\cite{With90} based on the Kondo exchange model that if 
the DOS of the conduction band vanishes linearly at the
Fermi energy, the Kondo effect is suppressed for small values of the 
coupling constant. However, for the special case where the band PH symmetry 
is present, the critical coupling strength is still found to be finite. 
At a first glance, our result seems to differ from the earlier prediction 
based on the Kondo spin model by Ingersent~\cite{Inge96} that for the PH 
symmetry case the Kondo effect is absent for all values of coupling. 
The discrepancy with Ref.~\cite{Inge96} may come from the facts 
that: (i) in the present model, the local PH symmetry is still broken due 
to the asymmetric Anderson impurity and/or the single-occupancy 
constraint; (ii) the existence 
of the Kondo effect is sensitive to the local character of conduction 
electrons near the impurity; and the magnetic property of the impurity 
and the quasiparticle properties of superconductors  
interplay with each other,  that is, the presence of the impurity 
enhances the LDOS of the $d$-wave superconductor (as 
shown below), which in turn screens the impurity spin.  

\begin{figure}[h]
\vskip 1.1truein
\centerline{\epsfxsize=3.0truein \epsfbox{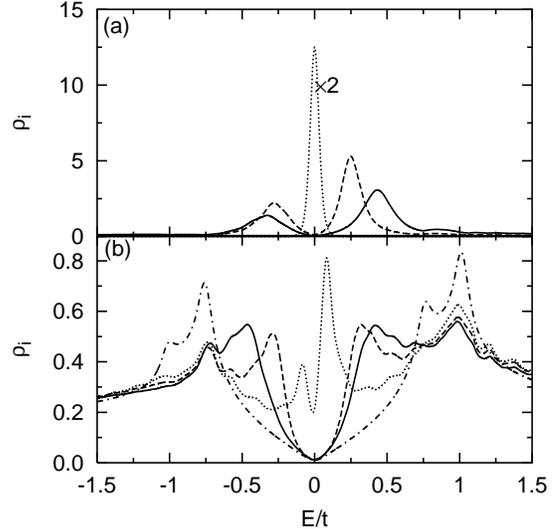}}
\vskip 0.1truein
\caption{The local density of states directly at the impurity site (a) 
and its nearest neighbor (b) for the coupling strength ${\cal V}=1t$ (solid 
line), $0.8t$ 
(dashed line), and $0.6t$ (dotted line). Also shown in (b) the bulk 
density of states (dash-dotted line). The chemical potential $\mu=-0.5t$.
}
\label{FIG:LDOS1}
\end{figure}

Once the parameter values of $\lambda_0$ and $b_0$, and the OP are 
obtained, the local density of states is then calculated according to 
\begin{equation}
\rho_{\bf i}=2\sum_{n}[\vert u_{\bf i}^{n}\vert^{2} \delta(E_{n}-E) 
+\vert v_{i}^{n} \vert^{2} \delta(E_{n}+E)]\;,
\end{equation}
which is proportional to the local differential tunneling conductance as 
measured by the STM. Figure 2 shows the LDOS as a function of energy 
at the impurity site and its nearest neighbor with $\mu=-0.5t$. 
Also shown in Fig.~\ref{FIG:LDOS1}(b) is the bulk density of states in 
the absence of the impurity. In the bulk DOS, the 
superconducting coherent peaks are exhibited at the gap edge 
$\pm \Delta_{g}(=0.75t)$. The peaks outside the gap comes from the Van 
Hove singularity. Due to the finite magnitude of the PH symmetry breaking 
by $\mu\neq 0$, the spectrum both inside and outside the gap is 
asymmetric with respect to $E=0$. More interestingly, within the energy 
gap, the resonant peaks show up in the LDOS both at the impurity 
site and its nearest neighbors, in sharp contrast to the unitary 
nonmagnetic impurity case where the LDOS vanishes at the impurity 
site~\cite{Salk96,Zhu00b}. 
The appearance of the double-peak structure in the LDOS directly at the 
impurity site also makes the result different from the weak nonmagnetic 
impurity case where only one peak appears below or above the Fermi energy 
($E=0$) when the scattering is repulsive or 
attractive~\cite{Salk96,Zhu00b}. The difference is mainly due to the fact 
that, in addition to a weak nonmagnetic-like impurity scattering in the 
Kondo resonance phase regime, there is also a disruption of the effective 
hopping as given by ${\cal V}b_0$ (from the point of view of the 
conduction electrons), immediately connecting the impurity site. The 
decrease of ${\cal V}$ enhances the quasiparticle scattering near the 
impurity by the suppression of ${\cal V}b_0$ on one hand and shifts the 
impurity level to $\tilde{\epsilon}_{d}=\epsilon_d+\lambda_0$
to the Fermi surface on

\begin{figure}[t]
\vskip 1.1truein
\centerline{\epsfxsize=3.0truein \epsfbox{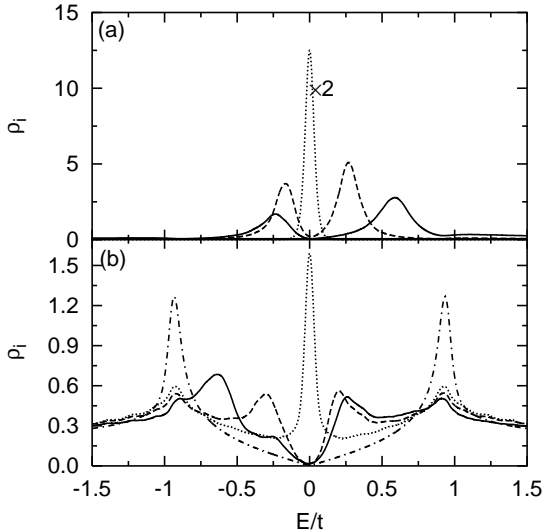} }
\caption{Same as Fig.~\ref{FIG:LDOS1} but with $\mu=0$.}
\label{FIG:LDOS2}
\end{figure}

\noindent
the other hand. As  a consequence of this decrease: (i) 
At the impurity site, the Kondo resonance peaks are sharpened with their 
separation decreased, and finally the double peaks merge into a 
single peak as ${\cal V}$ approaches ${\cal V}_{c}$ at which the 
decoupling of the impurity from the conduction band takes place; (ii) At 
the nearest neighbor to the impurity site, the separation between double 
peaks is also decreased but the double-peak structure still survives 
when ${\cal V}={\cal V}_{c}$, which is intrinsic to a $d$-wave 
superconductor with a short coherence when the PH symmetry 
is broken to some extent~\cite{Note3}. When the band PH symmetry is 
conserved ($\mu=0$), as shown in Fig.~\ref{FIG:LDOS2}, we see that most 
features of the local electronic structure are similar to the case with 
$\mu\neq 0$ except that the spectrum  is now symmetric outside the gap 
(but still asymmetric inside the gap due to the breaking of the local PH 
symmetry) and that at 
the nearest neighbor to the impurity only exhibits a single zero-energy 
peak when ${\cal V}={\cal V}_{c}$. We have also studied the case with 
different values of impurity level $\epsilon_d$ and found similar 
results. 

Two remarks are in order: (1) Although the slave-boson mean field theory
captures the essential physics in the Kondo resonance regime, where a
spin-singlet state is formed, unfortunately, it is unable to identify the
quantum transition from the spin-singlet state (screened impurity spin) 
to the spin-doublet state (unscreened impurity spin). The slave-boson
mean filed solution always leads to a lower energy in the spin-singlet
state than in the spin-doublet state~\cite{Rozh99}. 
One must go beyond the static mean-field level and employ 
alternatively the non-crossing approximation to describe this 
transition~\cite{Cler99}. (2) Very recently, Polkovnikov {\em et 
al.}~\cite{Polk00}  have studied a similar problem. 
The difference and similarity  between two works are as follows. 
The difference: (i) In their work, the Kondo exchange model where
the impurity spin is coupled to the spin of conduction electrons. These
two models should be equivalent, but the methods to solve them are
different;  (ii) In addition to the first nearest-neighbor hopping ($t$),
second ($t^{\prime}$) and the third ($t^{\prime\prime}$) nearest-neighbor
hopping were also introduced in their model.  Instead only the first
nearest-neighbor hopping is used in the present work; (iii) In their
work, the modification of the superconductivity by the Kondo impurity was
neglected, while  in the present work, this effect has been taken into
account by our self-consistent calculation; (iv) A vanishing density of
states near zero bias at the sites nearest neighboring to the impurity was
obtained in their work, while in the present work, the intensity
of density of states on these sites was found to be significant. The
similarity: (i) Both works have shown
the split peaks for the local density of states directly on the impurity
site; (ii) The models in both papers are placing the magnetic moment
at the impurity site. This is different from the
experimental situation for high-Tc superconductors with Zn impurities
where the magnetic spin would most likely be induced at 
sites nearest-neighboring to the Zn impurity.
 Therefore, these two models might be  more
suitable to describe the effects of Kondo impurities such as Mn, Fe, 
Co substituted for Cu in high-$T_c$ cuprates.

In summary, we have considered the problem of an Anderson impurity 
coupled to the conduction electrons of a $d$-wave superconductor.  
Within a slave-boson mean-field approach, we have been able to exactly 
diagonalize the composite Hamiltonian with the pair breaking effect taken 
into account. A critical coupling strength, above which the Kondo phase 
occurs, is found regardless of whether the PH symmetry is present. 
In the Kondo effect regime, the double-peak structure appears at both the 
impurity site and its nearest neighboring sites. 
Our analysis seems to  indicate that the spatial pattern of the
differential tunneling conductance around a Zn impurity in BSCCO~\cite{Pan00}
cannot be fully explained with the Kondo impurity model.

{\bf Acknowledgments}: We thank G. Kotliar, S. Sachdev, and P. Schwab for
useful discussions. This work was supported by 
the Texas Center for Superconductivity at the University of Houston and 
the Robert A. Welch Foundation.

\end{document}